\documentclass[oupdraft]{bio}
\usepackage{url}

\begin{document}

% Title of paper
\title{Developing a predictive signature for two trial endpoints using the cross-validated risk scores method}

% List of authors, with corresponding author marked by asterisk
\author{SVETLANA CHERLIN$^\ast$\\[4pt]
% Author addresses
\textit{Population Health Sciences Institute,
Newcastle University,
Baddiley-Clark Building, 
Newcastle upon Tyne,
UK}
\\[2pt]
% E-mail address for correspondence
{svetlana.cherlin@newcastle.ac.uk}\\[4pt]
JAMES M. S. WASON\\[4pt]
% Author addresses
\textit{Population Health Sciences Institute,
Newcastle University,
Baddiley-Clark Building, 
Newcastle upon Tyne,
UK \\and \\MRC Biostatistics Unit, University of Cambridge, Cambridge, UK}
\\[2pt]
}

% Running headers of paper:
\markboth%
% First field is the short list of authors
{S. Cherlin and J. M. S. Wason}
% Second field is the short title of the paper
{Cross-validated risk scores method for two trial endpoints}

\maketitle

% Add a footnote for the corresponding author if one has been
% identified in the author list
\footnotetext{To whom correspondence should be addressed.}

\begin{abstract}
{The existing cross-validated risk scores (CVRS) design has been proposed for developing and testing the efficacy of a treatment in a high-efficacy patient group (the sensitive group) using high-dimensional data (such as genetic data). The design is based on computing a risk score for each patient and dividing them into clusters using a non-parametric clustering procedure. In some settings it is desirable to consider the trade-off between two outcomes, such as efficacy and toxicity, or cost and effectiveness. With this motivation, we extend the CVRS design (CVRS2) to consider two outcomes. The design employs bivariate risk scores that are divided into clusters. We assess the properties of the CVRS2 using simulated data and illustrate its application on a randomised psychiatry trial. We show that CVRS2 is able to reliably identify the sensitive group (the group for which the new treatment provides benefit on both outcomes) in the simulated data. We apply the CVRS2 design to a psychology clinical trial that had offender status and substance use status as two outcomes and collected a large number of baseline covariates. The CVRS2 design yields a significant treatment effect for both outcomes, while the CVRS approach identified a significant effect for the offender status only after pre-filtering the covariates.}
{Clinical Trials; High-dimensional data; Innovative design; Multiple outcomes; Precision medicine; Risk scores.}
\end{abstract}

\section{Introduction}
\label{intro}

It is common in clinical trials that only a subgroup of treated patients is likely to benefit from an experimental therapy (\citealp{rothenberg:etal:2003, foster:etal:2011, zhao:etal:2013, janes:etal:2017}). This creates the need to better design and analyse clinical trials so they provide more information about which patients, if any, benefit from a treatment. For this purpose, high-dimensional information that is  increasingly being collected on patients can be used. There have been several approaches proposed for utilising (potentially high-dimensional) treatment by covariate interactions to stratify patients. For example, \citet{freidlin:simon:2005} and \citet{freidlin:etal:2010} proposed an adaptive signature design (ASD) that  combines a prospective development of a sensitive patient classifier and validation of the classifier in a single trial, based on treatment-covariate interactions obtained via regression modelling. Motivated by the work of \citet{freidlin:simon:2005} and \citet{freidlin:etal:2010}, \citet{zhang:etal:2017} focus on subgroup selection using baseline covariates by incorporating a utility function that takes into account the size of a subgroup or the possibility for an alternative treatment for the patients. In the randomised clinical trial of evaluation of the effects of the dose of dialysis and the level of flux of the dialyser membrane on time to death from any cause (\citealp{eknoyan:etal:2002}), the method identified a subgroup with a significantly greater treatment difference. Another approach to handling treatment-covariate interactions is decision tree algorithms (\citealp{su:etal:2009, dusseldorp:etal:2010, lipkovich:etal:2011}). These algorithms recursively partition the data with splits chosen to optimize an objective function. The algorithms differ in the size of the search space (\citealp{lipkovich:etal:2011}), the order of the splitting variables  \citep{dusseldorp:etal:2010}, the selection of covariates and the choice of the optimal cut-off (\citealp{loh:2002, loh:etal:2015}), and preventing bias in the variable selection by fitting a parametric model to the data in each node of the tree in a model-based recursive partitioning approach (\citealp {hothorn:etal:2006, zeileis:etal:2008,  seibold:etal:2016, krzykalla:etal:2020}). ``Virtual twins'' (\citealp{foster:etal:2011}) is another tree-based approach that involves computing the difference between predicted response probabilities for the treatment subject and its control ``twin''.  \citet{tian:etal:2014} proposed estimating treatment by covariate interactions using a modified covariate approach without the need for modeling the main effects. \citet{shen:he:2015}  tested the existence of subgroups with differential treatment effects by utilising the association between the subgroup membership and subject-specific covariates in a logistic-normal mixture model framework. A few studies have considered the compound covariate predictor approach for the prediction of predefined tumour classes (\citealp{radmacher:etal:2002}), for quantitatively estimating treatment effects and for predicting survival curves (\citealp{matsui:etal:2012, matsui:2006}). In the compound covariate predictor approach, the compound covariate was constructed using a test statistic from treatment-covariate interactions. \citet{zhao:etal:2013} selected the subgroups via a variety of  parametric scoring systems that were constructed as a function of multiple baseline covariates and were used to estimate the treatment difference. \citet{rosenwald:etal:2002} utilised gene-expression profiles to identify subgroups of patients with different survival rates, while the subgroups with distinctive gene-expression profiles were defined on the basis of hierarchical clustering.

All these methods have been proposed for a single endpoint.  However, in many clinical trials, multiple outcomes are of interest. This is especially relevant for clinical trials that analyse both efficacy and toxicity of the treatment (common in oncology and in other areas where treatment can have considerable side effects), efficacy and quality of life, and cost effectiveness of new drugs and  treatments in health economics. For example, \citet {zhao:leblanc:2019} considered the appropriate subpopulation for a clinical trial which was defined by a single biomarker that exceeded a specified threshold value. In addition to the main outcome that represented a response to the treatment, they proposed to incorporate other outcomes such as risk or cost associated with a new treatment by optimising  population impact. The  cross-validated risk scores (CVRS) method (\citealp{cherlin:wason:2020}) is based on constructing risk scores from a large number of baseline covariates. Risk scores represent a scoring system that is developed to be associated with the treatment effect and that can be used for predicting a benefit from the treatment. The CVRS design consists of two steps. In the first step, a single-response regression model is fit to every covariate and the risk scores are constructed as sums of associated covariates within each patient weighted by their estimated effects. In the second step, the risk scores are divided into two clusters that correspond to sensitive and non-sensitive groups of patients. Similarly to the cross-validated ASD (\citealp{freidlin:etal:2010}), the sensitive group is the subgroup of patients predicted to have high treatment effect. The method has been implemented in an R package \textit {rapids} (\citealp{rapids}). In this article, we propose the extension of the CVRS method that considers two outcomes, i.e.\ a design that develops a signature from a large number of covariates, considering two endpoints. The proposed CVRS2 method utilises  vector generalized linear models (\citealp {yee:2010}) for constructing bivariate risk scores. The bivariate risk scores are then divided into a prespecified number of clusters using an extension of the clustering procedure that has been used for the CVRS method. This clustering procedure is computationally straightforward and can be used as a starting point (we discuss other clustering procedures in the Discussion section). We explore the operating characteristics of the CVRS2 method for simulation scenarios that assume two outcomes that have different correlation structures. We also have extended the R package  \textit {rapids}  to incorporate the CVRS2 method. To our knowledge, CVRS2 is the first method within the adaptive signature design family that considers two outcomes. The remainder of this paper is organised as follows. The Methods section briefly describes the original CVRS design and introduces the CVRS2 design. In the Results section, we explore the operating characteristics of the CVRS2 design for various simulation scenarios. In the Real Data Example section, we illustrate the application of the CVRS2 design  to a randomised psychiatry clinical trial. Finally, we summarise our conclusions in the Discussion section.

\section{Methods}
\label{methods}

\subsection{CVRS Design} \label{sec:cvrs}

The full details of the method are provided in \citealp{cherlin:wason:2020}. Briefly, the model assumes that  the response to treatment denoted by $Y$ is influenced by a subset of $K$ unknown covariates (the sensitive covariates) though a generalised linear model. For example, for a binary outcome,  the model looks as follows: $\mathrm{log} \bigg ( \frac{p_i}{1-p_i} \bigg ) = \mu + \lambda t_i + \alpha_1 x_{i1} + \dots + \alpha_K x_{iK} + \gamma_1 t_i x_{i1} + \dots + \gamma_K t_i x_{iK}$,
where $p_i$ is the probability of the response to treatment for the $i$th patient; $\mu$ is the intercept; $\lambda$ is the treatment main effect that all patients experience regardless of the values of the covariates; $t_i$ is the treatment that the $i$th patient receives ($t_i$ = 0 for the control arm and $t_i$ = 1 for the treatment arm); $x_{i1}, \dots, x_{iK}$ are the values for the $K$ unknown sensitive covariates; $\alpha_1, \dots, \alpha_K$ are main covariate effects for the $K$ covariates; $\gamma_1, \dots, \gamma_K$ are the treatment-covariate interaction effects for the $K$ covariates. The model assumes that there is a subset of patients (the sensitive group) with a higher probability of response when treated with the new treatment compared with the control treatment. In the first step of the CVRS design, a signature is developed by constructing risk scores within each patient as sums of associated covariates weighted by their estimated effects. In the second step, a clustering procedure is applied to divide the risk scores into two clusters that correspond to sensitive and non-sensitive groups of patients. For constructing the risk scores, the cross-validation procedure is used, in which the model is built using the training subset and the risk scores are constructed for the patients in the test subset set as follows. For $r$-fold cross-validation, the observed dataset $D$ of size $N$ is randomly divided into $r$ non-overlapping subsets $D^{(l)}$, $l=1, \dots, r$, of (approximately) equal size $N/r$. A common choice of $r$ is 10 which we adopt here.  For each iteration of the $r$-fold cross-validation, data are split into test $D^{(l)}$ and training $D^{(-l)}$ (formed by removing $D^{(l)}$ from $D$) subsets and the coefficients for treatment by covariate interaction $\hat\beta_j^{(-l)}$ are estimated from $\mathrm{log} \bigg ( \frac{p_i}{1-p_i} \bigg ) = \mu + \lambda t_i + \alpha_j x_{ij} +\beta_j t_i x_{ij}$, for each covariate $j$ using training subset alone. Then, for each test set $D^{(l)}$, the risk scores are computed as $\mathrm{RS}_i^{(l)} = \sum_j \hat\beta_j^{(-l)} x_{ij}^{(l)}$ , where $x_{ij}^{(l)}$ is the value of  the covariate $j$ for the $i$th patient in the $l$th test set. Within each test set  $D^{(l)}$, the $k$-means procedure (\citealp{hartigan_wong:1979}) with $k$ = 2 is applied to classify the test scores $\mathrm{RS}_i^{(l)}$, $i = 1, \dots, N/r$ into sensitive and non-sensitive groups. Therefore, at the end of the cross-validation process each patient in the observed data $D$ is classified either as sensitive or non-sensitive, after pooling group membership status across the $r$ test sets. There are a few ways that one can test for the difference between the arms. One way is to consider the overall test positive if there is either a significant difference in the overall comparison between the arms or a significant difference between the arms in the sensitive group. The test for the overall comparison between the arms could be performed using a test for the difference of two proportions, carried out at a significance level $\alpha_1$, while for the comparison between the arms within the sensitive subgroup only Fisher's exact test could be carried out at a significance level $\alpha_2$.  The overall type I error is controlled at the significance level $\alpha = \alpha_1  + \alpha_2$. Alternatively, one can test for the interaction effect between the treatment and the sensitivity status using a generalised linear model. Generally, it is advisable to use a permutation method to obtain a $P$-value for testing the interaction effect between the treatment and the sensitivity status (\citealp{simon:etal:2004}) because when the sensitive group is obtained by cross-validation, the samples are not independent and therefore the standard asymptotic theory does not apply. In the permutation method, the entire cross-validation procedure is performed for every permuted data set and the corresponding test statistic is obtained. The one-sided permutation $P$-value is given by $\frac{1 + number~of~elements~of~\boldsymbol P^* \leq P_0}{1+number~of~permutations}$, where $\boldsymbol P^*$ is the vector of the $P$-values for the treatment-sensitivity status interaction effect computed  for a large number of permuted data sets, and $P_0$ is the $P$-value for the treatment-sensitivity status interaction effect obtained for the original (non-permuted) data.

\subsection{CVRS2 Design} \label{sec:cvrs2}

The CVRS2 design considers two outcomes, $Y_1$ and $Y_2$ (e.g.\ these could represent efficacy and toxicity in cancer clinical trials), that are influenced by a subset of $K$ unknown covariates (the sensitive covariates) through a vector generalized linear model (\citealp {yee:2010}). For example, the relationships between two binary outcomes and sensitive covariates can be described through a bivariate odds ratio model (\citealp {yee:2015}) as follows: 
	$\mathrm{log} \bigg ( \frac{p^{(1)}_i}{1-p^{(1)}_i} \bigg ) = \mu^{(1)} + \lambda^{(1)} t_i + \alpha^{(1)}_1 x_{i1} + \dots + \alpha^{(1)}_K x_{iK} + \gamma^{(1)}_1 t_i x_{i1} + \dots + \gamma^{(1)}_K t_i x_{iK}; ~
	\mathrm{log} \bigg ( \frac{p^{(2)}_i}{1-p^{(2)}_i} \bigg ) = \mu^{(2)} + \lambda^{(2)} t_i + \alpha^{(2)}_1 x_{i1} + \dots + \alpha^{(2)}_K x_{iK} + \gamma^{(2)}_1 t_i x_{i1} + \dots + \gamma^{(2)}_K t_i x_{iK}; ~
	\mathrm{log} (\psi_i) = \frac{\frac{p^{(1)}_i}{1-p^{(1)}_i}|p^{(2)}_i = 1}{\frac{p^{(1)}_i}{1-p^{(1)}_i}|p^{(2)}_i = 0}$,
where $p^{(1)}_i$ is the probability of the first outcome ($Y_1$) for the $i$th patient, $p^{(2)}_i$ is the probability of the second outcome ($Y_2$) for the $i$th patients, $\psi$ is the  odds ratio (ratio of odds of $Y_1 = 1$ when $Y_2 = 1$ and odds of $Y_1 = 1$ when $Y_2 = 0$). A measure that describes the association between the two proportions, $\psi$, is a natural measure for the association	between two binary responses. It is modelled as intercept-only, which is equivalent to the assumption of constant correlation. The responses are independent if and only if $\psi = 1$. A more complex  modelling of $\psi$ that allows for more flexible correlation structures can be used. However in some cases these may lead to numerical problems (\citealp{yee:2015}). Here, the estimated correlation between the outcomes is not explicitly used in the analysis. However, taking the correlation into account by jointly modelling the outcomes allows us to better estimate the effects of the covariates. For constructing the risk scores, the coefficients for treatment by covariate interaction are estimated for each covariate $j$ from a single-covariate bivariate odds ratio model:
$\mathrm{log} \bigg ( \frac{p^{(1)}_i}{1-p^{(1)}_i} \bigg ) = \mu^{(1)} + \lambda^{(1)} t_i + \alpha^{(1)}_j x_{ij} +\beta^{(1)}_j t_i x_{ij}; ~
\mathrm{log} \bigg ( \frac{p^{(2)}_i}{1-p^{(2)}_i} \bigg ) = \mu^{(2)} + \lambda^{(2)} t_i + \alpha^{(2)}_j x_{ij} +\beta^{(2)}_j t_i x_{ij}; ~
\mathrm{log} (\psi_i) = \frac{\frac{p^{(1)}_i}{1-p^{(1)}_i}|p^{(2)}_i = 1}{\frac{p^{(1)}_i}{1-p^{(1)}_i}|p^{(2)}_i = 0}$.
The cross-validation procedure is used for obtaining the bivariate risk scores which are computed for each patient $i$ as $\mathrm{RS}_i^{(l)}$ = $\{ \sum\limits_j \hat\beta^{(-l,1)}_j x_{ij}^{(l)}, \sum\limits_j \hat\beta^{(-l,2)}_j x_{ij}^{(l)}\}$ for each iteration of the $r$-fold cross-validation procedure.
Here, the first score is computed with respect to the first outcome, while the second score is computed with respect to the second outcome. The bivariate risk scores can be partitioned into a number of clusters within each test set  $D^{(l)}$. In some settings, e.g. where a treatment could be either (i) safe and effective; (ii) not safe and not effective; (iii) safe and not effective; (iv) effective and not safe, the most natural choice for the number of clusters is four. Letting the rate of the first response represent the  efficacy of the treatment and the rate of the second response represent the safety of the treatment, the four clusters would  represent (i) participants that have high response rates for both outcomes; (ii) participants that have low response rates for both outcomes, (iii) and (iv)  participants who have a high response rate for one of the outcomes and a low response rate for the other. This division can be accomplished by applying the $k$-means clustering procedure with $k$ = 4. Here, a sensitive group can be defined as one of the clusters or a combination of the clusters. In settings where a treatment would have to be either sufficiently safe and effective or not safe and not effective (i.e.\ when there are only two underlying clusters of patients), it may make better sense to divide patients into two groups. Here, the first group would represent participants with high response rates for both outcomes, while the second group would represent participants with low response rates for both outcomes. In this case, the $k$-means is applied with $k$ = 2. We note that $k$-means clustering with $k$ = 2 may not necessarily divide patients into two clusters with this interpretation. We explore this in the results and consider alternatives in the discussion. For each outcome, the power to conclude treatment effect in the trial population and the power to conclude treatment effect in the sensitive group are computed. We also consider the overall power to reach at least one of these conclusions. For the power in the trial population,  we compute the probability to detect a significant effect in the first outcome and  the probability to detect a significant effect in the second outcome, i.e.\ $\boldsymbol{P}^{tp} = \{P^{tp}_i\}$ where $i = 1, 2$ is the outcome.
In the four cluster case, for each outcome, we compute the power for the sensitive group as a set of four elements, each element  represents the probability to detect a significant difference between the treatment and the control in each one of the clusters, \i.e. $\boldsymbol{P}^{sens}_i = \{P^{sens}_{i1}, P^{sens}_{i2}, P^{sens}_{i3}, P^{sens}_{i4}\}$, where $i = 1,2$ is the outcome.
The overall power, $\boldsymbol{P}^{ov}_i =  \{P^{ov}_{i1}, P^{ov}_{i2}, P^{ov}_{i3}, P^{ov}_{i4}\}$, which represents the probability of a significant result for either the trial population or the sensitive group, is computed as follows: $\boldsymbol{P}^{ov}_i =  \{P^{ov}_{i1}, P^{ov}_{i2}, P^{ov}_{i3}, P^{ov}_{i4}\} = 
\boldsymbol{P}^{tp} + (\boldsymbol{1}-\boldsymbol{P}^{tp}) \times \boldsymbol{P}^{sens}_i =  
\{P^{tp}_i  + (1-P^{tp}_i) \times P^{sens}_{i1}, P^{tp}_i  + (1-P^{tp}_i) \times P^{sens}_{i2}, 
P^{tp}_i  + (1-P^{tp}_i) \times P^{sens}_{i3}, P^{tp}_i  + (1-P^{tp}_i) \times P^{sens}_{i4}\}.$
To compute the cluster-wise sensitivity and specificity in the four cluster case, we assume that each cluster in turn corresponds to a sensitive group. A detailed explanation of computing sensitivity and specificity is given in the ``Computing sensitivity and specificity" section of the Supplementary Materials.

\subsection{Simulation study}  \label{sec:sim}

We conducted a simulation study to evaluate the performance of the CVRS2 for four scenarios, exploring a different number of subgroups and the correlation between the covariates. For every scenario, we assumed a clinical trial with 100 covariates where $K$=10 of them are sensitive for each outcome with an overlap of five sensitive covariates, i.e. there were five covariates that were sensitive to both outcomes. The main effects of the covariates were assumed to be 0, and the treatment-covariate interaction effects were assumed to be constant across the sensitive covariates, similarly to \citealp{cherlin:wason:2020}: $\mathrm{log} \bigg ( \frac{p^{(1)}_i}{1-p^{(1)}_i} \bigg ) = \mu^{(1)} + \gamma^{(1)}_1 t_i x_{i1} + \dots + \gamma^{(1)}_K t_i x_{iK};~
\mathrm{log} \bigg ( \frac{p^{(2)}_i}{1-p^{(2)}_i} \bigg ) = \mu^{(2)} + \gamma^{(2)}_1 t_i x_{i1} + \dots + \gamma^{(2)}_K t_i x_{iK}$.
The intercepts $\mu^{(1)}$ and $\mu^{(2)}$ were set to correspond to control arm response rates of 25\%. We used an overall significance level  $\alpha$ = 0.05 (two-sided) that corresponds to $\alpha_1$ = 0.04 and $\alpha_2$ = 0.01 significance levels for the trial population test and for the sensitive group test, respectively, as suggested in \cite{freidlin:simon:2005}. The empirical overall power was calculated as the percentage of replications with either a positive trial population 0.04 level test or a positive 0.01 level sensitive group test. The results of the simulations were based on 1000 replications. The parameters used in the simulation study are presented in Table \ref{tab:Table1}. In all of the scenarios, the response rates and the sample sizes correspond to those used in \cite{freidlin:simon:2005} and \cite{cherlin:wason:2020}. A detailed description of how data were simulated is given in the ``Simulation steps" section of the  Supplementary Materials. To construct the risk scores, we fitted bivariate odds ratio model without the main treatment effect and the main covariate effects, to match the data generating mechanism:
	$\mathrm{log} \bigg ( \frac{p^{(1)}_i}{1-p^{(1)}_i} \bigg ) = \mu^{(1)} + \beta^{(1)}_j t_i x_{ij}; ~ 
	\mathrm{log} \bigg ( \frac{p^{(2)}_i}{1-p^{(2)}_i} \bigg ) = \mu^{(2)} + \beta^{(2)}_j t_i x_{ij}; ~ 
	\mathrm{log} (\psi_i) = \frac{\frac{p^{(1)}_i}{1-p^{(1)}_i}|p^{(2)}_i = 1}{\frac{p^{(1)}_i}{1-p^{(1)}_i}|p^{(2)}_i = 0}.$
We note that there is no direct comparator method that we are aware of for the two outcome case, hence we are comparing the CVRS2 method to the original CVRS method applied separately to each one of the outcomes, for both the simulations and the real data example. To assess the sensitivity of the CVRS2 method to various model misspecifications and extreme values of the parameters, we performed a sensitivity analysis (see the ``Sensitivity analysis" section of the Supplementary Materials).

\section{Results} \label{sec:res}
\subsection {Scenario I}  \label{sec:sc1}
The data were simulated assuming that there was a group of patients that had high rates of both responses $Y_1$ and $Y_2$ (the sensitive group), while the rest of the patients had low rates of both responses (the non-sensitive patients). Here, by response rate we denote the probability that the response takes the value of 1. The rates of $Y_1$ and $Y_2$ were 0.7 and 0.25 for the sensitive and the non-sensitive group, respectively (the response rates are illustrated in  Figure \ref{fig:Fig1}(a) for 10 simulations runs). The percentage of patients in the sensitive group were either 10\% or 20\%, the sample size was 400. The data were simulated assuming an independence between the covariates. The results of the analysis with a model that employs $k$ = 2 clusters (to match the data generating mechanism) are presented in Table \ref{tab:Table2}. The resultant risk scores are illustrated in Figure \ref{fig:Fig2}(a) for 10 simulation runs. The risk scores show a perfect separation between the sensitive and the non-sensitive groups, and the response rates in both sensitive and non-sensitive groups are estimated with a high precision(estimated rates of $Y_1$ and $Y_2$ for the sensitive and non-sensitive group were 0.69 and 0.25, respectively). 

\subsection {Scenario II}  \label{sec:sc2}
The data were simulated assuming four clusters of patients, which represent (i) patients with low response rates for both responses (cluster 1); (ii) and (iii) patients with a high response rate for one of the responses and a low response rate for the other, and vice versa (clusters 2 and 3); and (iv) patients with high response rates for both responses (cluster 4), as explained in Section \ref{sec:cvrs2}.  This scenario covers three sub-scenarios, IIa, IIb and IIc. In all three of the sub-scenarios the low response rate is 25\%, while the high response rate is 80\% for scenario IIa, 70\% for scenario IIb and 60\% for scenario IIc.  The response rates are illustrated in Figure \ref{fig:Fig1}(b-d) for 10 simulation runs. The mean rates of $Y_1$ and $Y_2$  for cluster 1 were 25\% for all three of the sub-scenarios. In Scenario IIa, the mean rate of $Y_1$ for clusters 3 and 4 was 80\%, and the mean rate of $Y_2$ for clusters 2 and 4 was 80\%. For Scenarios IIb and IIc, these mean response rates were 70\% and 60\%, respectively (see Table \ref{tab:Table1}). The covariates were assumed to be independent. The results are presented in Table \ref{tab:Table3} for sample sizes 400 and 1000. For each scenario we investigated the method assuming that the sensitive group corresponds to one cluster in turn, as elaborated in Section \ref{sec:cvrs2}. The risk scores for Scenario IIa are illustrated in Figure \ref{fig:Fig2}(b) for 10 simulation runs. The sensitivity and specificity of identifying the sensitive group are high reaching values $>$ 0.95 in many cases. Interestingly, the response rates are better estimated for clusters 1 and 4 rather than for clusters 2 and 3. This is in line with the results for Scenario I, suggesting that the method estimates the sensitive group better for clusters where the rates of both responses are similar. Very low power for the sensitive group test when cluster 1 corresponds to sensitive group suggests that the type I error is well controlled: considering cluster 1 as the sensitive group is equivalent to a null scenario.

\subsection {Scenario III}  \label{sec:sc3}
\begin{figure} [!p]
\centering\includegraphics[scale=0.9]{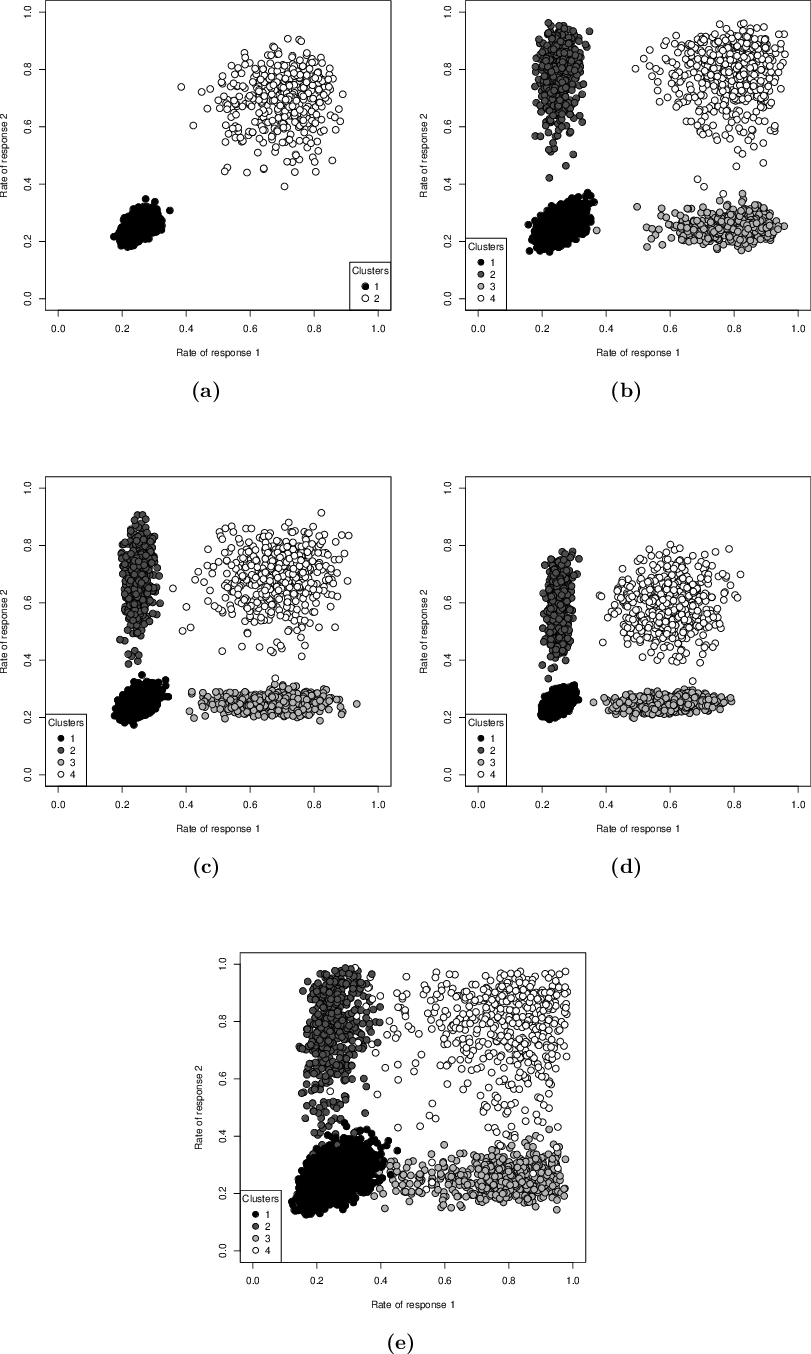}
\caption{The response rates on treatment for the simulated data for (a) scenario I, (b) scenario IIa, (c) scenario IIb, (d) scenario IIc, (e) scenario III.}
\label{fig:Fig1}
\end{figure}

\begin{figure} [!p]
\centering\includegraphics[scale=0.9]{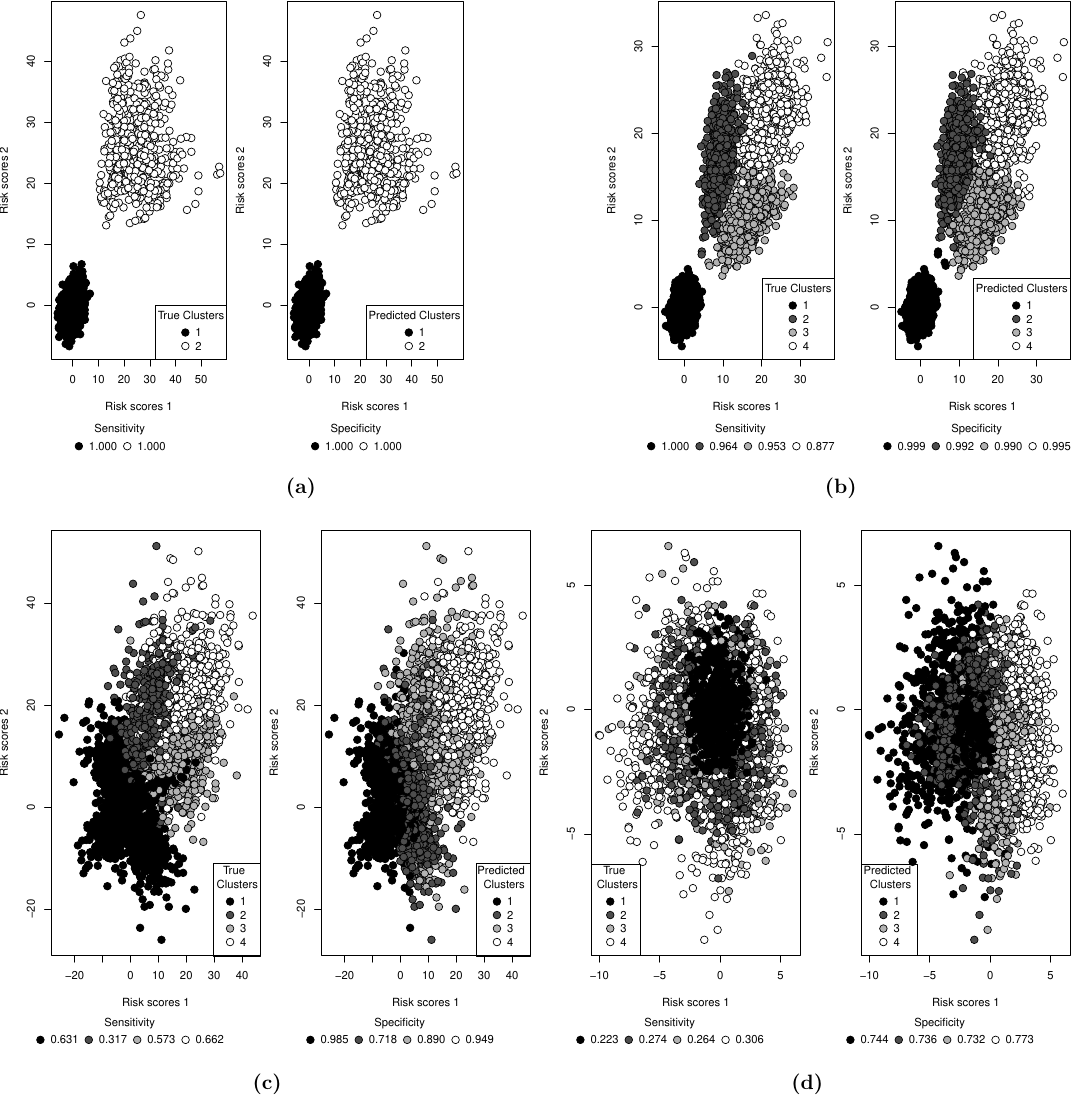}
\caption{The risk scores for (a) scenario I, (b) scenario IIa, (c) scenario III, (d) scenario IV.}
\label{fig:Fig2}
\end{figure}

\begin{figure} [!p]
	\centering\includegraphics[scale=0.85]{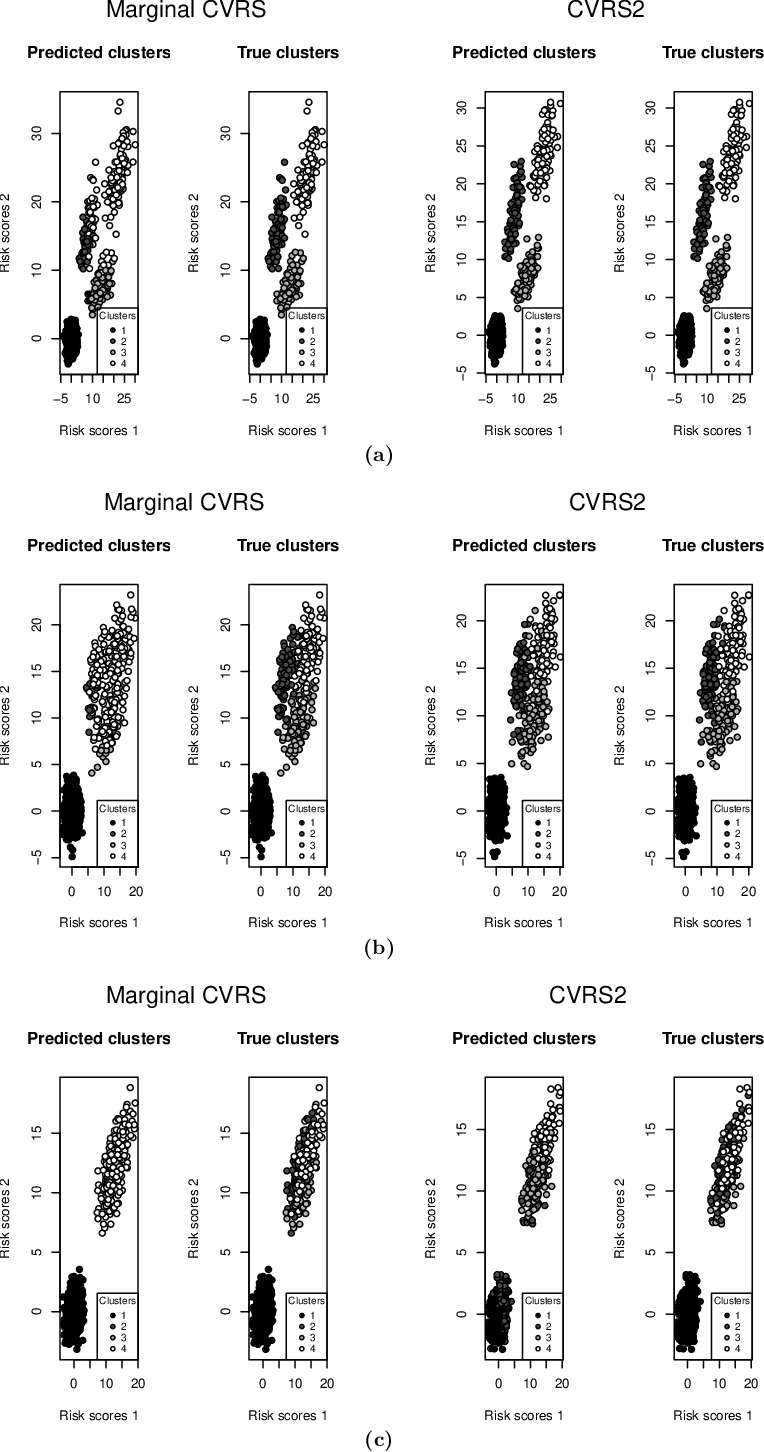}
	\caption{The risk scores for the marginal CVRS and the CVRS2 for the scenarios with (a) 0 overlapping covariates; (b) 5 overlapping covariates; (c) 9 overlapping covariates.}
	\label{fig:Fig3}
\end{figure}

To investigate the sensitivity of the CVRS2 method to the data generating mechanism, we simulated data assuming a moderate pairwise correlation ($\rho$ = 0.4) between all covariates. In this scenario, the data were simulated assuming four clusters of patients, with the low response rate being 25\% and the high response rate being 80\%, similarly to scenario IIa. The only difference with scenario IIa is the correlation between the covariates. The response rates are illustrated in Figure  {\ref{fig:Fig1}(e) for 10 simulation runs. Clearly, there is an increased variability of the cluster-wise response rates in comparison to scenarios with independent covariates (see Figure \ref{fig:Fig1}(b) for comparison). The results are presented in Table \ref{tab:Table3}. The risk scores are illustrated in Figure \ref{fig:Fig2}(c) for 10 simulation runs. Lower sensitivity and specificity values in comparison to scenario IIa (where $\rho$ = 0) suggest that the method would benefit from pre-filtering of the covariates based on the correlation between them. However, even without modelling the correlation between the covariates the method was still able to increase the overall power beyond that achieved for the trial population 0.04 level test.

\subsection {Scenario IV}  \label{sec:sc4}
The data were simulated assuming a null scenario where the rates of both responses $Y_1$ and $Y_2$ in all patients on both arms are 25\%. The results are presented in Table \ref{tab:Table3}. The risk scores are illustrated in Figure \ref{fig:Fig2}(d) for 10 simulation runs. The results show that a type I error is well controlled at the 0.01 level for the sensitive group test and at the 0.05 level overall. The response rates are estimated with high precision, being very close to 0.25 for all clusters (see Table \ref{tab:Table3}).

\subsection {Comparison between CVRS2 and CVRS}\label{sec:comparison}

In order to  compare the performance of the CVRS2 with the CVRS, we applied the CVRS to each one of the responses separately (we refer to this method as marginal CVRS). For the marginal CVRS, patients that were non-sensitive to both outcomes were referred to as cluster 1, patients that where sensitive to one of the outcomes were referred to as clusters 2 and 3, and patients that where sensitive to both outcomes were referred to as cluster 4. To illustrate the difference in the results between the CVRS2 and the marginal CVRS, we used three randomly selected simulation replicates with different correlations between the response rates. The correlation between the response rate was induced by the covariates that affect both responses (the overlapping covariates). All three of the data sets were simulated similarly to scenario IIb, i.e.\ the high response rate was 0.7, the sample size was 1000, the number of sensitive covariates was 10 for either response. The numbers of the overlapping covariates were zero, five and nine leading to estimated correlations of} -0.003, 0.24 and 0.46, respectively. We applied the CVRS2 and the marginal CVRS to each data set. The results as shown by the risk scores coloured by the predicted and true clusters are illustrated in Figure \ref{fig:Fig3}(a) for zero overlapping covariates, Figure \ref{fig:Fig3}(b) for five overlapping covariates and Figure \ref{fig:Fig3}(c) for nine overlapping covariates. A detailed description on how the marginal CVRS assigns the risk scores to four cluster is given in the ``Assignment of the risk scores to four clusters by marginal CVRS" section of the Supplementary Materials. In the case of zero overlapping covariates, the CVRS2 perfectly separated the clusters, while the marginal CVRS did not perform as well. In the case of five overlapping covariates, the CVRS2 had a much better ability to classify the patients in comparison to the marginal CVRS that failed to separate clusters 2, 3 and 4. However, in the case of nine overlapping covariates both designs were unable to classify the patients correctly into four clusters. This can be explained by the fact that a higher correlation between the response rates makes the data more compatible with two clusters than four, as illustrated by the risk scores. Interestingly, the marginal CVRS successfully differentiated between the two clusters of the data.
When we  applied the CVRS2 with $k$ = 2 to this data set, the design was able to perfectly separate the two underlying clusters, similarly to the marginal CVRS.

\section{Real Data Example}  \label{sec:real}

Previously, we have applied the CVRS method to the data from the Systematic Therapy of At Risk Teens (START) (\citealp{fonagy:etal:2018}). START was a randomised controlled trial comparing the outcomes of young people and their families who were allocated to treatment as usual (control arm) and multisystemic therapy (treatment arm). The data set comprised of 669 participants (336 participants in the control arm and 333 participants in the treatment arm) and 86 covariates. Participants with one or more offences were defined as offenders (288 in total; 143  in the control arm and 145 in the treatment arm), while participants with no offences were defined as non-offenders (381 in total; 193 in the control arm and 188 in the treatment arm). No overall significant treatment effect as measured by a logistic regression was detected ($P$ = 0.797).  
The CVRS method indicated the existence of a sensitive group comprised of 453 participants (222 participants in the control arm and 231 participants in the treatment arm) with no significant interaction effect between the treatment and the sensitivity status (permutation-based $P$ = 0.122). However, a marginally significant interaction effect between the treatment and the sensitivity status ($P$ = 0.043) was achieved after a pre-filtering of the covariates based on a $P$-value threshold. Here, we analyse the START data considering two outcomes: in addition to the previously analysed offender/non-offender status, we analyse a binary substance use status (the numbers of the offenders/substance users are presented in Supplementary Table 1). Due to missingness of the substance use status in above 30\% of participants, the two outcome dataset comprises of 461 participants (218 participants in the control arm and 243 participants in the treatment arm). The analysis was performed without the pre-filtering of the covariates. No overall significant treatment effect as measured by a bivariate odds ratio model was detected ($P$ = 0.993 and $P$ = 0.181 for the offender status and the substance use status, respectively). We analysed the START data with the CVRS2 method assuming either two or four underlying clusters  (see Supplementary Figure 1 for the corresponding risk scores). For constructing the risk scores, we fitted the bivariate odds ratio model that includes the main treatment effect and the main covariate effects, as described in Section \ref{sec:cvrs2}. For the two-cluster analysis, the  CVRS2 method found a sensitive group comprised of 283 participants (132 participants in the control arm and 151 participants in the treatment arm).  The permutation-based $P$ values for the interaction between the treatment and the sensitivity status were 0.003 and 0.129 with respect to the offender status and the substance use status, respectively (based on 2000 permutations). 
The permutation-based $P$-value was computed assuming one cluster in turn corresponds to a sensitive group. With respect to the offender status, the cluster-wise permutation $P$-values were 0.007, 0.147, 0.023 and 0.162, while with respect to the substance use, the cluster-wise permutation $P$-values were 0.038, 0.193, 0.509 and 0.299. The estimated cluster-wise rates for the offender status were 0.27, 0.36, 0.44 and 0.49, while the estimated cluster-wise rates for the substance use status were 0.23, 0.4, 0.36 and 0.43. The size of the clusters were 67, 132, 151, and 111 subjects. The mean offender rate and substance use rate in each arm within each cluster are shown in Supplementary Table 3. In this study it would make sense to assume that cluster 1 that corresponds to low offence rate and low substance use rate, represents the sensitive group. Our results indicate that there is a significant $P$-value of the interaction between the sensitivity status and the treatment with respect to both of the outcomes with respect to cluster 1. This suggests that while no treatment effect was indicated for the overall population of the participants, a subgroup of participants who belong in cluster 1 benefited from the treatment. In order to compare the CVRS2 with the CVRS, we applied the marginal CVRS to the two outcome dataset that comprises of 461 patients. The permutation-based $P$-values for the interaction effect between the treatment and the sensitivity status were $P = 0.101$ and $P = 0.383$ with respect to the offender status and the substance use status, respectively (Supplementary Figure 2). The coefficients of the covariates that contributed to the risk scores are presented in Supplementary Table 4 for the CVRS2 method and in Supplementary Table 5 for the marginal CVRS method. One could look at which covariates contribute most of the risk scores by multiplying the values of the covariates by the corresponding coefficients.

\section{Discussion} \label{sec:disc}

We have proposed a method that can stratify patients into groups that have different treatment effects on more than one outcome.  This method represents a modification of the CVRS design. The new CVRS2 design considers two outcomes and utilises bivariate risk scores. The scores are constructed as sums of the covariates weighted by their coefficients, estimated with vector generalised linear models. The method assumes four pre-defined clusters, based on the rate of the outcomes (see the ``Interpretation of the clusters" section of the Supplementary Materials for the interpretation of the clusters). We have  investigated the performance of the CVRS2 method by applying it to various simulated scenarios. We found that when the data is clearly divided into two subgroups of samples (sensitive and non-sensitive), the method is able to perfectly separate between the groups. When the data consists of four clusters (two clusters with both high/low rates of responses, and two clusters with a high rate of one of the responses and a low rate of the other), the method is able to identify the clusters and to estimate the rates of the responses reasonably well. We showed that in the case of four clusters, increasing the sample size, as well as increasing the response rates of a cluster with the highest response rate, improves the performance of the method. We investigated the correlation between the response rates in the sensitive groups on treatment by simulating the data with different numbers of overlapping covariates (zero, five and nine). We found that in the cases of zero and five overlapping covariates, the CVRS2 design performs better than the marginal CVRS design (the original one outcome CVRS that has been applied to each outcome separately). However, when the number of the overlapping covariates is too large (nine out of ten in our case), both methods are unable to identify four clusters, because the data are more compatible with two clusters. This suggests that the method is sensitive to the prior assumption about the number of the true clusters.

To illustrate the applicability of the methods to the real data, we have applied it to the data from a START randomised controlled trial. Here we considered the offender status and the substance use status as two outcomes. We showed that the CVRS2 was able to identify a sensitive group that conferred a significant interaction effect ($P$ = 0.003) between the treatment and the sensitivity status (with respect to the offender status), assuming the data consist of two groups of participants (interestingly, the interaction effect between the treatment and the sensitivity status with respect to substance use was not significant: $P$ = 0.129). This is in contrast to the CVRS method that considers one outcome only, which did not identify a sensitive group that conferred a significant interaction effect between the treatment and the sensitivity status. This shows that incorporating information from an additional outcome could marginally improve the results for the first outcome, even though the results for the second outcome are not significant. We next analysed the START data assuming there are four clusters of participants (see the ``Interpretation of the clusters" section of the Supplementary Materials for the interpretation of the clusters). The results showed a significant interaction effect between the treatment and the sensitivity status with respect to the offender status when cluster 1 (a cluster of patiens with low offence rate and low substance use rate) is assumed to belong to a sensitive group ($P$ = 0.007). The corresponding $P$-value for the substance use achieved a nominal significance as well  ($P$ = 0.038). The estimated response rates for the cluster with the highest response rates (0.49 and 0.42)  were lower than those of the simulated data. This suggests that for the real data with higher response rates for the sensitive group, the method would achieve more significant results. Additionally, a larger data set would achieve a higher power as illustrated for the simulated data. We compared the results for the CVRS2 design with the results for the marginal CVRS for the same set of patients, to investigate the importance of joint modelling of the outcomes. We showed that the marginal CVRS which is equivalent to independent modelling of the outcomes, was not able to find a significant interaction effect between the treatment and the sensitivity status with respect to both offender status and substance use status. This result is in line with the result for the simulated data that showed a better classification ability for the CVRS2 in comparison to the marginal CVRS. Interestingly, the risk scores for the CVRS2 and for the marginal CVRS are driven by different covariates, as can be seen from Supplementary Tables 4 and 5. The covariates that have the largest absolute values of the coefficients for the marginal CVRS, have coefficients equal to zero for the CVRS2 (for example, a diagnosis of eating disorder). The difference in the contribution of the covariates to the risk scores in the two methods could explain the difference in the assignment to the clusters. To further investigate the differences in the results of the CVRS2 and the marginal CVRS methods, we compared the values of the measure of association, $\psi$, for the real data for each covariate and for the simulated data (Scenario IIb, averaged over 100 simulations, for 100 covariates). Supplementary Figure 10 shows that the values of $\psi$ for the real data are consistently higher than one. This suggests that the two responses in the case study are not independent which could also explain the differences in the results from applying the CVRS2 and the marginal CVRS methods.

The membership in a cluster is specified by a combination of the baseline covariates. We note that while the clinical interpretation of the covariates is beyond the scope of the manuscript, it could facilitate further independent investigation. The risk scores might be utilised in designing  adaptive enrichment trials where only the patients who are predicted to benefit from the treatment are enrolled into the trial.
We note that CVRS method was able to achieve a significant result for the real data only after a pre-filtering of the covariates based on a $P$-value threshold. Here, by taking an additional outcome into account, the CVRS2 method achieved a significant result without applying any pre-filtering of the covariates, even for a smaller sample size.  However, the simulation results show that when the covariates are correlated, the method might benefit from the pre-filtering of the covariates based on the correlation between them, which will be investigated in future work. An additional issue that requires further investigation is the number of clusters. We have investigated the cases of two and four clusters. Yet, for the four-cluster case, one might want to investigate merging some of the clusters together into a sensitive group. In practice, the assumption of the number of the clusters might also depend on the nature of the study and the question to be analysed. Finding the optimal number of clusters would be a part of future research. Similarly to the CVRS method, in this study, we have retrospectively applied the CVRS2 method to identify the sensitive group in psychiatry trial participants. In principle, the method can be used to prospectively identify whether a participant belongs to a sensitive group. To divide the patients into clusters, we utilised a computationally straightforward and scalable $k$-means clustering procedure that works well in practice (\citealp{jian:2010}). For example,  $k$-means clustering has been used previously  in the context of gene expressions, where it successfully separated the high-risk versus low-risk cutoff for the up-/down-regulated mean ratio of gene expressions (\citealp{shaughnessy:etal:2007}). However, we note that different clustering algorithms can be utilised. For instance, \citet{rosenwald:etal:2002} utilised hierarchical clustering to analyse genes whose expression was correlated with the outcomes. In this study, we applied a supervised clustering approach, i.e.\ the number of clusters was pre-specified. This approach requires prior knowledge about the most plausible number of clusters, as shown in Section \ref{sec:comparison}. Future work will investigate the performance of the unsupervised clustering approaches,  where the number of clusters  is unknown \textit{a priori}, and on the sensitivity of the method to different clustering approaches. Additional areas to explore are the parametric clustering approach in which  each cluster is assumed to follow a parametric distribution and  the data is modelled with a mixture model (\citealp{fraley:raftery:2002}), and the semiparametric clustering approach that models high-density data with a parametric density and low-density data with a non-parametric density (\citealp{pan:etal:2019}). 

We note that our method gives equal weight to both endpoints. However in clinical practice the outcomes could have different importance for different patients. There are a few possible ways to address this issue. First, patients could interpret their pair of risk scores for the two outcomes, and weigh which treatment is preferable according to their relative importance of the two outcomes. This would allow for patients to apply their own judgement regarding the importance of the outcomes. Secondly, a univariate risk score could be formed that gives a pre-specified weight to each component of the bivariate risk score, i.e. $RS_i = w_1RS_{1i} + w_2RS_{2i}$, where $w_1$ and $w_2$ are the weights, and $RS_{1i}$ and $RS_{2i}$ are the components of the bivariate risk score $RS_i$ for patient $i$. This could then be included within the CVRS2 approach. Finally, a parametric type of clustering could incorporate the relative importance of each outcome into the clustering procedure.

Similarly to the CVRS method, the CVRS2 method is developed for binary outcome. It is straightforward to extend the implementation of the method to incorporate different types of outcomes, such as normally distributed  or time-to-event endpoints. Also, a more flexible joint model (e.g. latent variable  model) could be employed to handle mixtures of outcomes.

\section{Software} \label{soft}

We extended the R package \textit {rapids}  (\citealp{rapids}) to incorporate the CVRS2 method. The package is available at \url {https://github.com/svetlanache/rapids}.

\section{Supplementary Materials} \label{sec6}

Supplementary Materials are available online at \url{http://biostatistics.oxfordjournals.org}.

\section*{Acknowledgments}

The authors wish to thank Professor Peter Fonagy and the trial team for providing the data. This work was supported by The Sir Bobby Robson Foundation. JMSW is funded by the UK Medical Research Council (MC\_UU\_00002/6 and MR/N028171/1).

{\it Conflict of Interest}: None declared.

\bibliographystyle{biorefs}
\bibliography{refs}

%Figure 1

%Figure 2

%Figure 3

%Table 1

\begin{table}[!p] 
\tblcaption{Parameters that are used in simulation study. Each row corresponds to a simulated scenario. 
Column $\theta_{i},\sigma^2_{i}, \rho_{i} $ corresponds to the mean, variance and correlation of multivariate normal distribution that was used to simulate sensitive  covariates in the subgroups that are sensitive to outcome  $i = 1, 2$.
Column $\nu_{i},\zeta^2_{i}, \kappa_{i} $ corresponds to the mean, standard deviation and correlation of multivariate normal distribution that was used to simulate sensitive covariates in the subgroups that are not sensitive to outcome  $i = 1, 2$.
Column $\eta,\xi^2, \tau $ corresponds to the mean, standard deviation and correlation of multivariate normal distribution that was used to simulate the rest of the covariates in all patients.
For each cluster of patients, the columns $RR_1$ and $RR_2$ correspond to the rates of response 1 and response 2 red for the sensitive group on the experimental arm, respectively. \label{tab:Table1}}
{\tabcolsep = 4.25pt
\centering
\begin{tabular}{@{}cccccccccccc@{}}
\tblhead{
& & & & \multicolumn{2}{c} {Cluster 1}  & \multicolumn{2}{c} {Cluster 2} & \multicolumn{2}{c} {Cluster 3}  & \multicolumn{2}{c} {Cluster 4}\\ \cline{5-12}
Scenario & $\theta_{i},\sigma^2_{i}, \rho_{i}$ & $\nu_{i},\zeta^2_{i}, \kappa_{i}$ & $\eta,\xi^2, \tau$ & $RR_1$ &$RR_2$  & $RR_1$ &$RR_2$ & $RR_1$ & $RR_2$  & $RR_1$ &$RR_2$}
I  & 1, 0.25, 0     & 0, 0.01, 0    & 0, 0.25, 0   & 0.25 & 0.25 & 0.7   & 0.7 & -     & -      & -    & -     \\
IIa & 1, 0.25, 0    & 0, 0.01, 0    & 0, 0.25, 0   & 0.25 & 0.25 & 0.25 & 0.8 & 0.8 & 0.25 & 0.8 & 0.8  \\
IIb & 1, 0.25, 0    & 0, 0.01, 0    & 0, 0.25, 0   & 0.25 & 0.25 & 0.25 & 0.7 & 0.7 & 0.25 & 0.7 & 0.7  \\
IIc & 1, 0.25, 0    & 0, 0.01, 0    & 0, 0.25, 0   & 0.25 & 0.25 & 0.25 & 0.6 & 0.6 & 0.25 & 0.6 & 0.6  \\
III & 1, 0.25, 0.4 & 0, 0.01, 0.4 & 0, 0.25, 0.4 & 0.25 & 0.25 & 0.25 & 0.8 & 0.8 & 0.25 & 0.8 & 0.8  \\
IV & 1, 0.25, 0    & 0, 0.01, 0    & 0, 0.25, 0   & 0.25 & 0.25 & 0.25 & 0.25 & 0.25 & 0.25 & 0.25 & 0.25 
\lastline
\end{tabular}}
\end{table}

%Table 2
 
\begin{table}[!p] 
\tblcaption{Operating characteristics of the CVRS2 methods for scenario I. The response rates on the control arm are 25\%.  10 covariates are sensitive to  response 1, 10 covariates are sensitive to response 2 with the overlap of 5 covariates. The results are based on 1000 simulations. \label{tab:Table2}}
{\tabcolsep = 4.25pt
\centering
\begin{tabular} {@{}lccc@{}}
\tblhead{
&\multicolumn{2}{@{}c@{}}{Sensitive group}  \\\cline{2-3} 
Operating characteristics & 10\%& 20\%} 
       Power in the trial population (w.r.t. response 1)  & 0.123 & 0.460  &   \\  
       Power in the trial population (w.r.t. response 2)  &  0.125 & 0.446  &    \\  
       Power in the sensitive group (w.r.t response 1)& 0.515 & 0.892 &   \\  
       Power in the sensitive group (w.r.t. response 2) & 0.469 & 0.906  &   \\  
       Overall power (w. r. t. response 1) & 0.571 & 0.934 &  \\  
       Overall power (w. r. t. response 2) & 0.523 & 0.941 &  \\  
       Sensitivity of the group selection & 0.999 & 1.000&  \\  
       Specificity of the group selection  & 0.999 & 1.000 & \\  
       Estimated rate of response 1 in sensitive group & 0.694 & 0.689 & \\ 
       Estimated rate of response 2 in sensitive group & 0.689 & 0.693 &\\
       Estimated rate of response 1 in non-sensitive group & 0.250 &  0.251 & \\ 
       Estimated rate of response 2 in non-sensitive group & 0.251 &  0.251 &
\lastline
\end{tabular}}
\end{table}

 %Table 3
 \begin{table}[!p] 
\tblcaption{Operating characteristics of the CVRS2 method for scenarios II, III and IV. The response rates on the control arm are 25\%.  10 covariates are sensitive to response 1, 10 covariates are sensitive to response 2 with the overlap of 5 covariates. Each one of the clusters 2, 3, 4 comprise of 10\% of patients. The results correspond to sample size 400. The results in the parentheses correspond to sample size 1000. 
The power for the tiral population 0.04 level test  w. r. t. response 1 and response 2, respectively is:  0.595(0.956) and 0.596(0.956) for scenario IIa;
0.427(0.835) and 0.43(0.842) for scenario IIb;
0.289(0.604) and 0.294(0.594) for scenarion IIc;
0.961 and 0.951 for scenario III;
0.037 and 0.044 for scenario IV. \label{tab:Table3}}
{\tabcolsep = 4.25pt
\centering
\begin{tabular}{@{}clccccc@{}}
\tblhead{
& &\multicolumn{4}{@{}c@{}}{Sensitive group corresponds to:}  \\\cline{3-6} \rotatebox{90}
{Scenario} & Operating characteristics & Cluster 1 & Cluster 2 & Cluster 3 & Cluster 4}
\multirow{8}{*}{IIa} &
       Power in the sensitive group (w. r. t. resp. 1)&  0.007(0.009) & 0.010(0.017) & 0.606(0.998) & 0.751(0.998) &\\   &
       Power in the sensitive group (w. r. t. resp. 2)&0.004(0.010) & 0.441(0.998) & 0.067(0.025) & 0.581(0.997) & \\  &
       Overall power (w. r. t. response 1) & 0.599(0.956) & 0.599(0.956) & 0.829(1.000) & 0.889(1.000) &\\  &
       Overall power (w. r. t. response 2) & 0.597(0.957) & 0.766(1.000) & 0.619(0.957) & 0.826(0.999) & \\  &
       Sensitivity of the group selection & 0.925(1.000)& 0.779(0.964) & 0.776(0.953) & 0.790(0.877)  &\\  &
       Specificity of the group selection  & 0.999(0.999) & 0.927(0.992) & 0.962(0.990) & 0.980(0.995) &\\ &
       Estimated rate of response 1&0.250(0.251) & 0.287(0.279) & 0.695(0.774) & 0.788(0.791) &\\ &
       Estimated rate of response 2 &0.250(0.252) & 0.597(0.775) & 0.402(0.283) & 0.719(0.786) &
\lastline
\multirow{8}{*}{IIb} &
       Power in the sensitive group (w. r. t. resp. 1)&0.008(0.004) &0.019(0.023)& 0.297(0.938)& 0.548(0.951)&\\  &
       Power in the sensitive group (w. r. t. resp. 2)&0.010 (0.011)&0.124(0.933) &0.107(0.029) &0.389(0.955)& \\ & 
       Overall power (w. r. t. response 1) &0.431(0.836)& 0.437(0.838)& 0.589(0.988)& 0.734(0.992) &\\  &
       Overall power (w. r. t. response 2) &0.435(0.846)& 0.498(0.988)& 0.489(0.845)& 0.663(0.991) & \\ &
       Sensitivity of the group selection &0.800(0.997)& 0.544(0.928)& 0.711(0.909)& 0.782(0.846)  &\\  &
       Specificity of the group selection  &0.999(0.999) & 0.830(0.986)& 0.937(0.984)& 0.970(0.993) &\\ &
       Estimated rate of response 1&0.248(0.251)& 0.274(0.282)& 0.558(0.671)& 0.687(0.698) &\\ &
       Estimated rate of response 2 &0.251(0.252)& 0.403(0.661)& 0.433(0.300)& 0.621(0.684) &
\lastline
       \multirow{8}{*}{IIc} &
       Power in the sensitive group (w. r. t. resp. 1)&0.005(0.006) &0.009(0.023) &0.115(0.699) &0.348(0.773)&\\  &
       Power in the sensitive group (w. r. t. resp. 2)&0.010(0.013) &0.023(0.543) &0.108(0.077) &0.224(0.729)& \\ & 
       Overall power (w. r. t. response 1) &0.292(0.607)&  0.296(0.612)& 0.364(0.871) &0.530(0.901) &\\  &
       Overall power (w. r. t. response 2) &0.301(0.600) &0.309(0.815) &0.365(0.627)& 0.444(0.882) & \\ &
       Sensitivity of the group selection & 0.658(0.951)& 0.310(0.805)& 0.639(0.845)& 0.760(0.821)  &\\  &
       Specificity of the group selection  &0.996(0.999)& 0.731(0.946)& 0.906(0.969)& 0.955(0.989) &\\ &
       Estimated rate of response 1&0.250(0.250) & 0.262(0.284) & 0.446(0.550)& 0.581(0.600) &\\ &
       Estimated rate of response 2 &0.253(0.252)& 0.298(0.504)& 0.428(0.325) & 0.527(0.588) &
\lastline
\multirow{8}{*}{III} &
       Power in the sensitive group (w. r. t. response 1)& 0.010 &0.018 &0.789 &0.990\\   &
       Power in the sensitive group (w. r. t. response 2)& 0.014& 0.201& 0.691& 0.907 & \\ &
       Overall power (w. r. t. response 1) &  0.955& 0.955& 0.990& 0.999 &\\  &
       Overall power (w. r. t. response 2) & 0.945& 0.953& 0.981& 0.994 & \\ &
       Sensitivity of the group selection & 0.631 &0.317 &0.573 &0.662  &\\  &
       Specificity of the group selection  & 0.985& 0.718& 0.890& 0.949 &\\ &
       Estimated rate of response 1 &0.250 &0.278 &0.540 &0.736 &\\ &
       Estimated rate of response 2 & 0.254& 0.343& 0.511& 0.657 &
\lastline
\multirow{8}{*}{IV} &
       Power in the sensitive group (w. r. t. response 1)&  0.012 & 0.011 & 0.008 & 0.010\\   &
       Power in the sensitive group (w. r. t. response 2)&0.006 & 0.004 & 0.012 & 0.009 & \\ &
       Overall power (w. r. t. response 1) & 0.048 & 0.047 & 0.044 & 0.047 &\\  &
       Overall power (w. r. t. response 2) & 0.050 & 0.047 & 0.055 & 0.052 & \\ &
       Sensitivity of the group selection & 0.223& 0.273 & 0.264 & 0.306  &\\  &
       Specificity of the group selection  & 0.744 & 0.736 & 0.732 & 0.773 &\\ &
       Estimated rate of response 1 &0.248 & 0.249 & 0.249 & 0.252 &\\ &
       Estimated rate of response 2 &0.252 & 0.248 & 0.251 & 0.249 &
\lastline
\end{tabular}}
\end{table}

\end{document}